\RequirePackage{fix-cm}
\documentclass{svjour3}

\smartqed  
\usepackage{amsmath} 
\usepackage{amssymb}  
\usepackage{graphics} 
\usepackage{graphicx}
\usepackage{epsfig} 
\usepackage{pstool}
\usepackage{eucal}
\usepackage{flushend}
\usepackage{upgreek}
\usepackage{tabularx}
\usepackage{booktabs} 
\usepackage{color}
\usepackage{lipsum}
\usepackage{flafter} 
\usepackage{stfloats} 
\usepackage{bm}
\usepackage{float}
\usepackage[american]{babel}
\usepackage{placeins}
\usepackage{siunitx}

\graphicspath{{figures/}}

\journalname{Int. J. Comput. Assist. Radiol. Surg.}

\usepackage{hyperref}

\begin{document}

\title{Intrapapillary Capillary Loop Classification in Magnification Endoscopy: Open Dataset and Baseline Methodology
\thanks{This work was supported through an Innovative Engineering for Health award by Wellcome Trust [WT101957]; Engineering and Physical Sciences Research Council (EPSRC) [NS/A00027/1] and a Wellcome / EPSRC Centre award [203145Z/16/Z \& NS/A000050/1].}
}

\titlerunning{A Benchmark Dataset and Evaluation Methodology for IPCL Classification} 

\author{Luis C. Garc\'ia-Peraza-Herrera \and
        Martin Everson \and
        Laurence Lovat \and
        Hsiu-Po Wang \and
        Wen Lun Wang \and
        Rehan Haidry \and
        Danail Stoyanov \and
        S\'ebastien Ourselin \and
        Tom Vercauteren
}

\authorrunning{Luis C. Garc\'ia-Peraza-Herrera et al.} 

\institute{   L. C. Garcia-Peraza-Herrera \at
              Department of Medical Physics and Biomedical Engineering, UCL, London, UK\\
              \email{luis.herrera.14@ucl.ac.uk}
            \and
              M. Everson, L. Lovat, R. Haidry \at
              Division of Surgery \& Interventional Science, UCL, Department of Gastroenterology; University College Hospital NHS Foundation Trust, London, UK
            \and
              H. Wang \at
              Department of Internal Medicine, National Taiwan University, Taipei,  Taiwan
            \and
              W. L. Wang \at
              Department of Internal Medicine, E-Da Hospital/I-Shou University, Kaohsiung, Taiwan
            \and
              D. Stoyanov \at
              Wellcome / EPSRC Centre for Interventional and Surgical Sciences, UCL, London, UK
            \and
             L. C. Garcia-Peraza-Herrera, S. Ourselin, T. Vercauteren \at
             School of Biomedical Engineering \& Imaging Science, KCL, UK
}

\date{}

\maketitle

\begin{abstract}~\\
%
\textbf{Purpose} 
Early squamous cell neoplasia (ESCN) in the oesophagus is a highly treatable condition. Lesions confined to the mucosal layer can be curatively treated endoscopically.
We build a computer-assisted detection (CADe) system that can classify still images or video frames as normal or abnormal with high diagnostic accuracy.\\
%
\textbf{Methods}
We present a new benchmark dataset containing $68$K binary labeled frames extracted from $114$ patient videos whose imaged areas have been resected and correlated to histopathology. Our novel convolutional network (CNN) architecture solves the binary classification task and \textit{explains} what features of the input domain drive the decision-making process of the network.\\
%
\textbf{Results} 
The proposed method achieved an average accuracy of \SI{91.7}{\percent}
compared to the 
\SI{94.7}{\percent}
achieved by a group of $12$ senior clinicians. 
Our novel network architecture produces deeply supervised activation heatmaps
that suggest the network is looking at intrapapillary capillary loop (IPCL) patterns when predicting abnormality.\\
\textbf{Conclusion}
We believe that this dataset and baseline method may serve as a reference for future benchmarks on both video frame classification and explainability in the context of ESCN detection. A future work path of high clinical relevance is the extension of the classification to ESCN types.


\end{abstract}
\section{Introduction}
\label{sec:introduction}
Oesophageal cancer is the sixth most common cause of cancer deaths worldwide~\cite{Zhang2013} and a burgeoning health issue in developing nations from Africa along a `cancer belt' to China. The current gold standard to investigate oesophageal cancer is gastroscopy with biopsies for histological analysis.
\begin{figure}
    \centering
    \includegraphics[width=\textwidth]{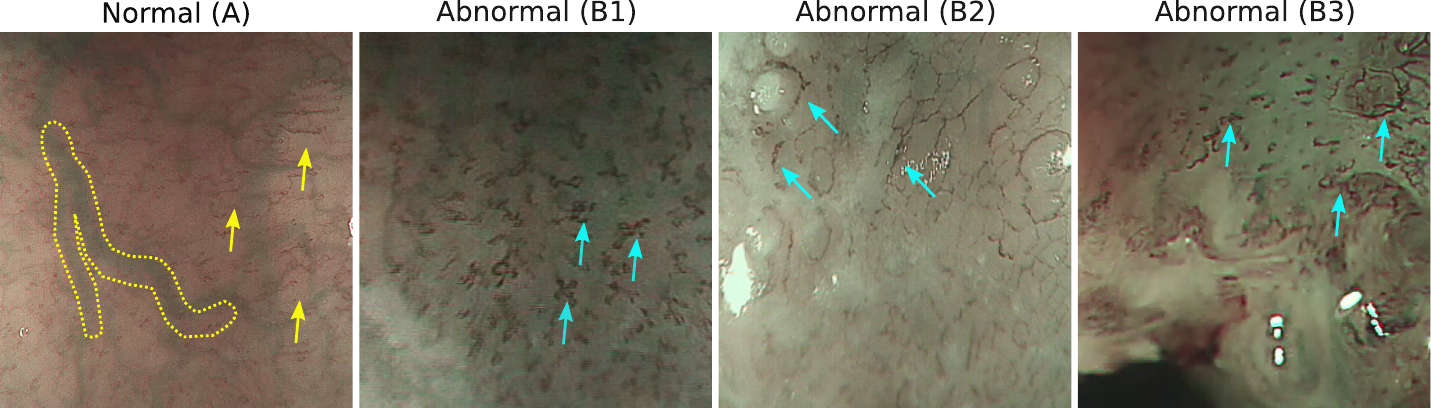}
    \caption{Magnifying endoscopy (ME) frames extracted from videos of patients with different histopathology. Normal patients typically present a clear deep submucosal vasculature, large green-like vessels such as the one highlighted within the dashed yellow line are usually visible. Intrapapilary capillary loops (IPCL) refers to the microvasculature (pointed by the arrows). Healthy patients tend to present thinner (yellow arrows) and less tangled IPCL patterns than those with abnormal tissue (blue arrows).}
    \label{fig:ipcl}
\end{figure}
Early squamous cell neoplasia (ESCN) is a highly treatable type of oesophageal cancer, with recent advances in endoscopic therapy meaning that lesions confined to the mucosal layer can be curatively resected endoscopically with a $<$\SI{2}{\percent} incidence of local lymph node metastasis \cite{Cho2014}. The endoscopic appearances of ESCN lesions are subtle and easily missed, with significant miss rates on endoscopy within the three years preceding diagnosis~\cite{Menon2014}. Early cancers invading into the submucosa are likely to have local lymph node metastasis and should be referred promptly for consideration of surgical resection.

Intrapapillary capillary loops (IPCL) are a clinical microvascular feature recognised as an endoscopic marker for ESCN~\cite{Inoue1996,Sato2015,Oyama2017}. They have been classified by the Japanese Endoscopic Society (JES) in a simplified system aimed at improving the easy recognition of ESCN by endoscopists \cite{Oyama2017}. The type of IPCLs present also facilitate the accurate prediction of the lesion histology; Type A IPCLs (see fig. \ref{fig:ipcl}) correlate with normal tissue. Type B1, B2, B3 IPCLs (see fig. \ref{fig:ipcl}) demonstrate progressive morphologic abnormalities and correlate with the invasion of early neoplasia in the muscularis mucosa and submucosal tissue. Oyama et al. \cite{Oyama2017} demonstrate the JES classification offers high diagnostic accuracy compared to other classifications for the prediction of dysplastic tissue – with the overall accuracy for histology prediction \SI{90.5}{\percent} across type B1-3.
%
%
A computer-assisted detection (CADe) system that can classify still images or video frames as normal or abnormal with high diagnostic accuracy could provide a useful adjunct to both expert and inexpert endoscopists. 

\subsection{Contributions}
We focus on the problem of classifying video frames as normal/abnormal. These frames are extracted from the magnification endoscopy (ME) recording of a patient. To the best of our knowledge, we introduce the first IPCL normal/abnormal open dataset\footnote{https://github.com/luiscarlosgph/ipcl} containing ME video sequences correlated with histopathology. Our dataset contains $68$K video frames from $114$ patients. 

For a small and representative sample of $158$ frames (IPCL types A, B1, B2, B3) we ask $12$ senior clinicians to label them as normal/abnormal and report the inter-rater agreement as Krippendorff's $\alpha$ coefficient \cite{Krippendorff2004}, achieving \SI{76.6}{\percent}. We also draw a comparison between raters and our \emph{gold standard} histopathology results, achieving an average accuracy across raters of \SI{94.7}{\percent}.

We propose a novel convolutional network (CNN) architecture to solve the binary classification task with a particular focus on the explainability of predictions. Our proposed method achieved an average accuracy of \SI{91.7}{\percent}. In addition to a global classification estimation, our novel design produces activation maps and class scores at every resolution of the convolutional pyramid. 
The network has to \textit{explain} where it is \textit{looking at} prior to the generation of a class prediction. Looking at the activation maps for the abnormal class, we have observed that the network is \textit{looking at} IPCL patterns when predicting abnormality. No conclusive evidence has been found that it is paying attention to large deep submucosal vessels to detect normal tissue. We believe that this baseline method may serve as a reference for future benchmarks on both video frame classification and explainability in the context of ESCN detection.


\section{Related work}
\label{sec:related_work}

Computer-aided endoscopic detection and diagnosis could offer an adjunct in the endoscopic assessment of ESCN lesions; there has been a high level of interest in recent years in developing clinically interpretable models. The use of CNNs has shown potential across several medical specialties. In gastroenterology, considerable efforts have been devoted to the detection of malignant colorectal polyps \cite{Wang2019,Hassan2019,Su2019} and upper gastrointestinal cancer \cite{Luo2019}. However, its utility in endoscopic diagnosis of early oesophageal neoplasia remains in its infancy~\cite{Everson2019}. 

Guo et al. \cite{Guo2019} propose a CNN that can classify images as dysplastic or non-dysplastic. Using a dataset of 6671 images, they demonstrate per frame sensitivity of \SI{98}{\percent} for the detection of ESCN. Using a video dataset of $20$ videos, they demonstrate per frame sensitivity of \SI{96}{\percent} for the detection of ESCN. Although the results are encouraging, the size of the patient sample is limited. Given the \textit{black box} nature of CNNs this may represent a matter of concern with regards to generalization capability. Zhao et al. \cite{Zhao2019} have also reported a CNN for the classification of IPCL patterns in order to identify ESCN. Using 1383 images, although heavily skewed towards Type B1 IPCLs, they demonstrated overall accuracies of \SI{87}{\percent} for the classification of IPCL patterns. In this study, however the authors excluded type B3 IPCLs from the training and testing phase. The CNN also demonstrated only a \SI{71}{\percent} classification rate for normal IPCLs, indicating that it over-diagnoses normal tissue as containing type B1 IPCLs, and so representing dysplastic tissue.

\section{Dataset details}
\label{sec:datasets}
This dataset will be made publicly available online upon publication and can thus serve as a benchmark for future work on detection of ESCN based on magnification endoscopy images.

\subsection{Patient recruitment, endoscopic procedures and video acquisition}
Patients attending for endoscopic assessment to two early squamous cell neoplasia (ESCN) referral centres in Taiwan (National Taiwan University Hospital and E-Da Hospital) were recruited with consent. Patients with oesophageal ulceration, active oesophageal bleeding or Barrett’s oesophagus were excluded. 
%
Gastroscopies were performed by two expert endoscopists (WLW, HPW), either under conscious sedation or local anaesthesia. An expert endoscopist was defined as a consultant gastroenterologist performing $>50$ Early Squamous Cell Neoplasia (ESCN) assessments per year. All endoscopies were performed using an HD ME-NBI GIF-H260Z endoscope, with Olympus Lucera CV-290 processor (Olympus, Tokyo, Japan). A solution of water of simethicone was applied via the endoscope working channel to the oesophageal mucosa, in order to remove mucus, food residue or blood. This allowed good visualization of the oesophageal mucosa and microvasculature, including IPCLs. 

\subsection{Correlating imaged areas with histology}
\label{sec:imaged_areas}
Initially, a macroscopic assessment was made of the suspected lesion in an overview, with the borders of the lesion delineated by the endoscopist. The endoscopist then identified areas within the borders of the lesion on which to undertake magnification endoscopy. The IPCL patterns were interrogated using magnification endoscopy in combination with narrow-band imaging (ME-NBI). Magnification endoscopy was performed on areas of interest at $80-100$x magnification. Using the JES IPCL classification system the IPCL patterns were classified by the consensus of three expert endoscopists (WW, HPW, RJH) as type A, B1, B2, B3, in order to give a prediction of the \textit{worst-case} histology for the whole lesion. The entire lesion was then resected by either endoscopic mucosal resection (EMR) or endoscopic submucosal dissection (ESD). Resected specimens were formalin-fixed and assessed by an expert gastrointestinal histopathologist. As is the gold standard the \textit{worst case} histology was reported for the lesion as a whole, based on pathological changes seen within the resected specimen. Similarly to abnormal lesion areas, type A recordings (normal, healthy patients) were obtained by visual identification of healthy areas, magnification endoscopy, visual confirmation of normal vasculature and IPCL patterns, and biopsy to confirm the assessment. 

\subsection{Dataset description}
Our IPCL dataset comprises a total of 114 patients ($45$ normal, $69$ abnormal). Every patient has a ME-NBI video ($30$fps) recorded following protocol in Section~\ref{sec:imaged_areas}. Raw videos can present some parts where NBI is active. In this dataset, only ME subsequences are considered. All frames are extracted and assigned to the class \textit{normal} or \textit{abnormal} depending on the histopathology of the patient. They are quality controlled one-by-one (running twice over all the frames) by a senior clinician with experience in the endoscopic imaging of oesophageal cancer. Frames that are highly degraded due to lighting artifacts (e.g. blur, flares and reflections) up to the point where it is not possible (for the senior clinician) to make a visual judgement of whether they are normal or abnormal are marked as uninformative and not used. 
This curation process results in a dataset of 67742 annotated frames ($28078$ normal, $39662$ abnormal) with an average of $593$ frames per patient. 
%
%
For each fold, patients (not frames) are randomly split into \SI{80}{\percent} training, \SI{10}{\percent} 
validation (used for hyperparameter tuning), and \SI{10}{\percent} testing (used for evaluation). 
The statistics of each individual fold are presented in the supplementary material.
%

%

\subsection{Evaluation per patient clip}
Let $\{\hat{y}_{f,p}\}_{f=1}^{F_p}$ be the set of estimated probabilities for the frames $f$ (out of $F_p$) belonging to patient clip $p$. Then, the estimated probability of abnormality for $p$ is computed as an average of frame probabilities:
\begin{equation}
    P\left(X=abnormal\Big|\{\hat{y}_{f,p}\}_{f=1}^{F_p}\right) = \frac{1}{F_p} \sum_{f=1}^{F_p} \hat{y}_{f, p}
    \label{eq:patient_estimation}
\end{equation}
Similarly to frame predictions, a threshold ($p=0.5$) is applied to obtain a class label for $p$. As per our data collection protocol (see \ref{sec:imaged_areas}), magnification endoscopy clips contain either normal or abnormal tissue. Hence, a correlation between $P(X=abnormal|\{\hat{y}_{f,p}\}_{f=1}^{F_p})$ and histopathology is expected. The analysis of clip classification errors facilitates the identification of worst cases, singling out patient-wide mistakes from negligible frame prediction errors.

\section{Methods}
\label{sec:methods}
\begin{figure*}
	\centering
	\includegraphics[width=\textwidth]{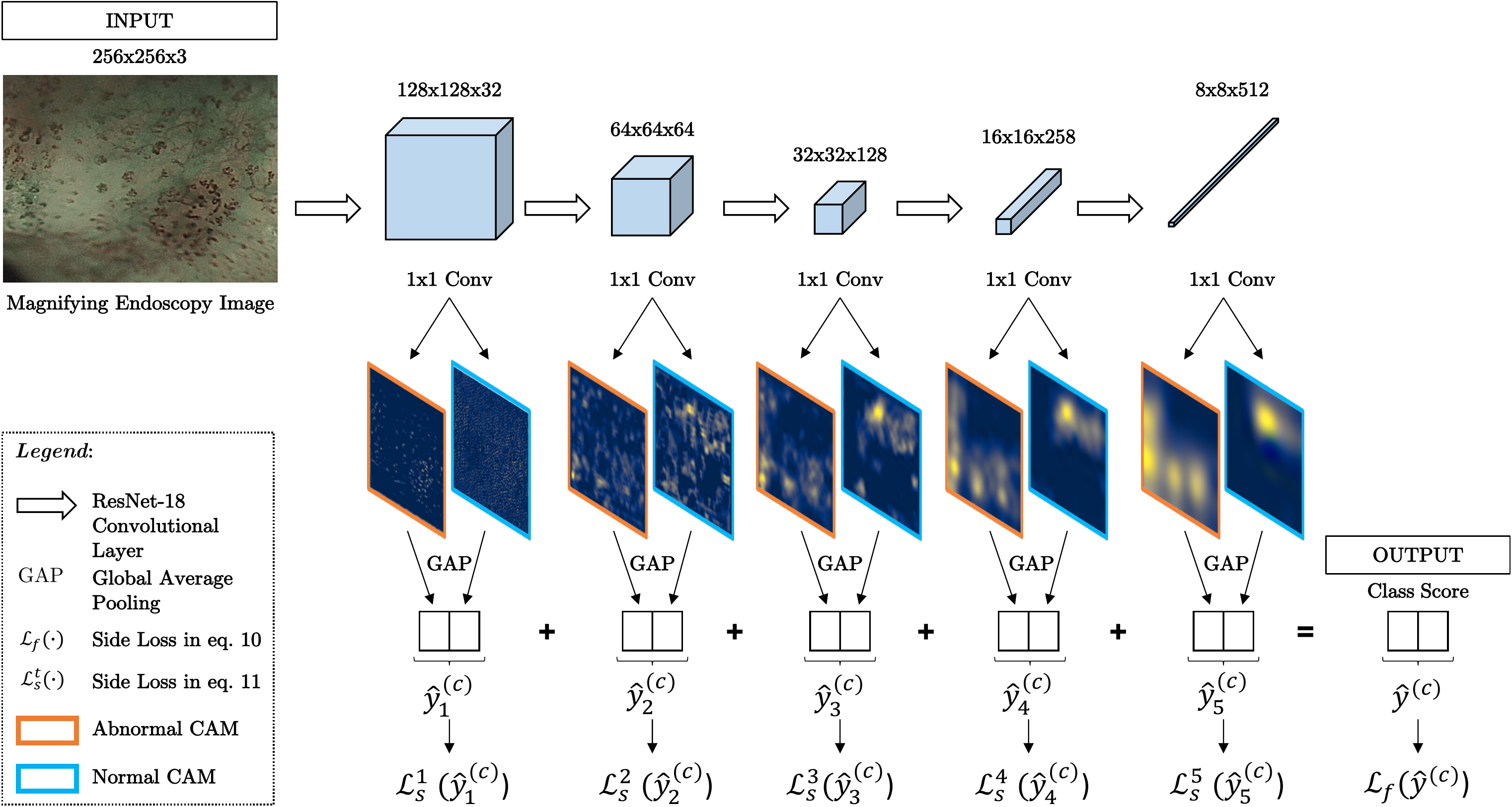}
	\caption{Proposed model ResNet-18-CAM-DS with embedded positive class activation maps at all resolutions.}
	\label{fig:cars_method}
\end{figure*}

In this section, we propose a reference method for IPCL binary classification with a particular focus on explainability that may serve as a baseline for future benchmarks.
%
%
As it is common in data-driven classification, we aim to solve for a mapping $f$ such that $f_{\boldsymbol{\theta}}(\boldsymbol{x}) \approx \boldsymbol{y}$, where $\boldsymbol{x}$ is an input image, $\boldsymbol{y}$ the class label corresponding to $\boldsymbol{x}$, and $\boldsymbol{\theta}$ a vector of parameters. 
All the input images were preprocessed by downscaling them to a width of $256$ pixels (height automatically computed from their original aspect ratio) so that we could fit a large batch of images into the GPU. To account for changes in viewpoint due to endoscope motion, random ($p=0.5$) on-the-fly flips are applied to each image. Our baseline model is
%
ResNet-18~\cite{He2016}. 
The batch normalization 
moving average fraction is set to $0.7$. Our batch size, momentum and weight decay hyperparameters are set to $256$, $0.9$, and $0.0005$ respectively. 
The initial learning rate (LR) 
was tuned by grid search. It was set to 
$\lambda=5e-3$ 
for training all folds, decaying it every $10$K iterations ($\approx40$ epochs) by a factor of $0.5$ until $45$K iterations ($\approx200$ epochs). 
In our implementation, 
using an NVIDIA GeForce TITAN X Pascal GPU and Caffe~$1.0$ as deep learning framework, the inference time per-frame is $7.6$ms [$6.4$ms, $9.9$ms], 
enabling the algorithm for deployment as a real-time endoscopy solution.
%

\subsection{Explaining network predictions, baseline without FC layer: ResNet-18-CAM}
Explaining network predictions is of particular interest to draw a comparison between image features that clinicians employ in their clinical practice and those that might exist but be unknown to them. Conversely, adding attention to those image features that are known to be relevant but are not used by the network could potentially improve its performance. In the context of ESCN detection, this leads to investigate whether the network is actually looking at deep submucosal vessels and IPCL patterns to predict abnormality. The answer to this question typically comes in the form of a heatmap, with those parts relevant to the classification being highlighted.

Our baseline model (ResNet-18) may be formalized as $f_{\boldsymbol{\theta}} = r(h(g(T_{\boldsymbol{x}})))$ where $T_{\boldsymbol{x}} = T_{\boldsymbol{\theta}}(\boldsymbol{x}) \in \mathbb{R}^{H \times W \times K}$ is the feature tensor obtained after processing $\boldsymbol{x}$ at the deepest pipeline resolution, $K$ represents the number of feature channels, $T_{\boldsymbol{x}}(k)$ is a matrix that represents the feature channel with index $k$, and $g$, $h$, and $r$ represent the global average pooling (GAP), fully connected (FC), and final scoring convolution layers respectively. 

The FC layer $h$ represents a challenge for explainability, as relevance is redistributed when gradients flow backwards, losing its spatial connection to the prediction being made \cite{Montavon2018}. Hence, inspired by \cite{Zhou2015}, we stripped out the fully connected layer of $1000$ neurons from the baseline model (ResNet-18), connecting the output of the GAP directly to the neurons that predict class score (those in layer $r$) and setting their bias to zero. 
Formally, this leads to $f_{\boldsymbol{\theta}} = r(g(T_{\boldsymbol{x}}))$, the output of the network before softmax being
\begin{equation}
    \boldsymbol{\hat{y}}^{(c)} = \sum_{k \in K} w_{k, c} \left[ \underbrace{\frac{1}{HW} \sum_{i, j}}_\text{GAP} \underbrace{T_{\boldsymbol{x}}(k)}_\text{Feature tensor} \right]
    \label{eq:resnet-18-cam-output}
\end{equation}
where $w_{k, c} \in \boldsymbol{\hat{\theta}}$, and $\boldsymbol{\hat{y}}^{(c)}$ is the score predicted for class~$c$. Following this approach, 
a heatmap per class can be generated obviating the GAP layer during inference, simply computing
\begin{equation}
    \boldsymbol{\hat{y}_{CAM}}^{(c)} = \sum_{k \in K} w_{k, c} T_{\boldsymbol{x}}(k)
    \label{eq:resnet-18-cam-heatmap}
\end{equation}
These heatmaps called Class Activation Maps (CAMs)~\cite{Zhou2015} keep a direct spatial relationship to the input, which is relevant for visual explanations.
Although the architecture proposed in \cite{Zhou2015} requires removing the GAP layer to produce the CAMs, \eqref{eq:resnet-18-cam-output} can be reformulated as 
\begin{equation}
    \boldsymbol{\hat{y}}^{(c)} = \underbrace{\frac{1}{HW} \sum_{i, j}}_\text{GAP} \underbrace{\left[ \sum_{k \in K} w_{k, c} T_{\boldsymbol{x}}(k) \right]}_\text{CAM}
    \label{eq:cam_reformulated}
\end{equation}
in which case the CAMs are embedded within the network pipeline as a $1\times1$ convolution (as we have already shown in \cite{Garcia-Peraza-Herrera2018}).
This leads to $f_{\boldsymbol{\theta}} = g(r(T_{\boldsymbol{x}}))$. We refer to this architecture as ResNet-18-CAM (as for the baseline, LR is set to $5e-3$ and decayed by $0.5$ every $10$K iterations until $45$K iterations). The performance of this network is shown in table \ref{tab:resnet_18_cam_result_frames}. Although the accuracy of ResNet-18-CAM is comparable to the baseline network (ResNet-18), ResNet-18-CAM conveniently computes a heatmap per class as part of the network processing. However, the explainability in the context of our classification problem remains very challenging due to the low resolution of the heatmaps produced. 

\subsection{Deeply Supervised Class Activation Maps: ResNet-18-CAM-DS}
In the computer vision field, 
images tend to display one or a few large objects. 
%
\begin{figure*}
	\centering
	\includegraphics[width=0.8\textwidth]{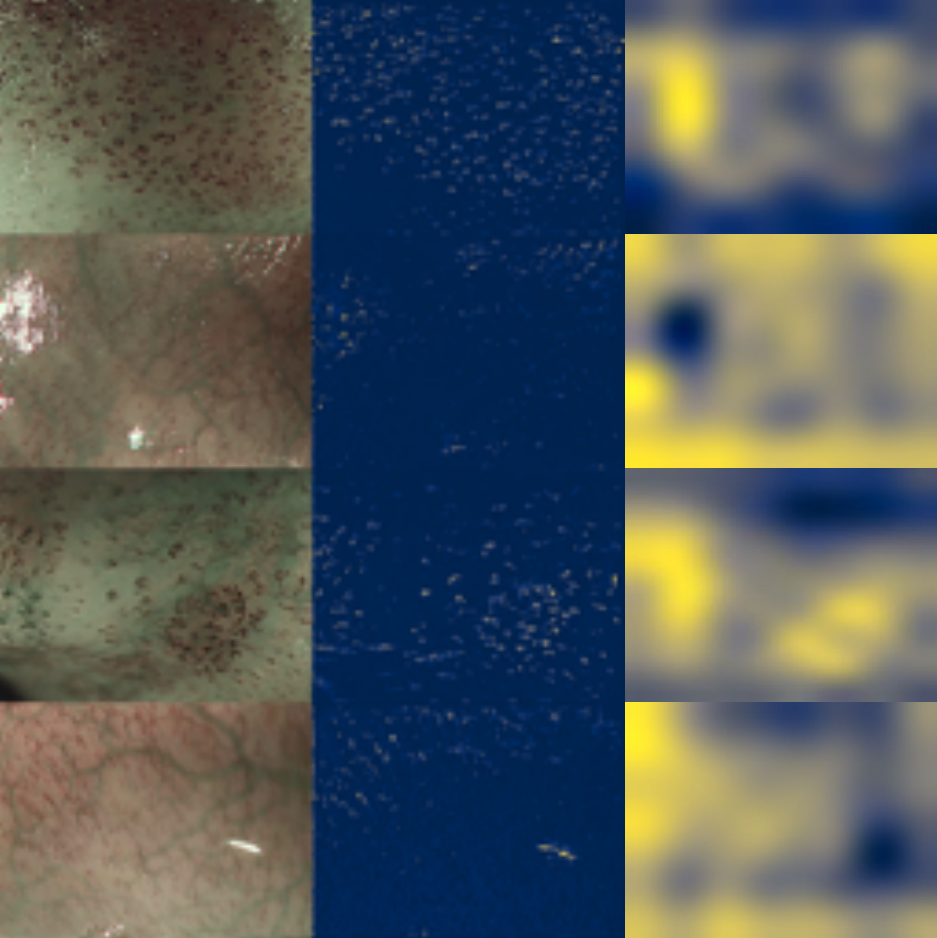}
	\caption{Representative images from the testing set of fold $1$ (left). Highest resolution CAM generated by ResNet-18-CAM-DS for the abnormal class (better viewed in the digital version). That is, $\boldsymbol{\hat{y}_{CAM_t}}^{(c)} = \boldsymbol{\hat{y}_{CAM_1}}^{(1)}$ (center). Class Activation Maps generated by ResNet-18-CAM \cite{Zhou2015} (right). In contrast to traditional CAMs generated by ResNet-18-CAM (right), ours (center) suggest that our network is \textit{looking at} IPCLs to predict abnormality.}
	\label{fig:ipcl_cams}
\end{figure*}
This is however, not the case in medical images such as the magnification endoscopy ones used to classify IPCL patterns. 
%
%
Due to their low resolution, it is very challenging to understand what the network is \textit{looking at}, as abnormal microvasculature in an endoscopic image is not localized only in a single spot.
%
In our clinical problem, two types of features could be expected to be highlighted, submucosal vessels and IPCLs, which represent endoscopic markers for ESCN~\cite{Inoue1996,Sato2015,Oyama2017}. The procedure to generate the CAM proposed in \cite{Zhou2015} employs the deepest feature maps as inputs to produce the attention heatmaps. For our input images of $256\times256$ pixels, these feature maps have a resolution of $8\times8$ pixels, leading to very low-resolution CAMs (also $8\times8$ pixels
). This hinders the explanatory capability of the heatmaps, as small capillaries are the main clinically discriminating feature. It is of interest to know whether they are being \textit{looked at} to predict abnormality. A trivial solution would be to reduce the depth of the network, but this could potentially hamper the learning of abstract features and decrease performance. In addition, the optimal amount of resolution levels for the given task to balance accuracy and explainability is a hyperparameter that would need to be tuned. Instead, we propose an alternative path modelling $f_{\boldsymbol{\theta}}(\boldsymbol{x})$ as
\begin{equation}
    f_{\boldsymbol{\theta}}(\boldsymbol{x}) = (
            f_{\boldsymbol{\theta_t}} 
            \circ 
            f_{\boldsymbol{\theta_{t-1}}}  
            \circ
            \cdots
            \circ 
            f_{\boldsymbol{\theta_2}} 
            \circ 
            f_{\boldsymbol{\theta_1}}
        )
        (\boldsymbol{x})
        \label{eq:proposed_model}
\end{equation}
where $f_{\boldsymbol{\theta_t}}$ represents the function that processes the input at resolution $t$, and whose output tensor has a width and height downsampled (strided convolution) by a factor of $0.5$ with regards to its input tensor. In this formulation, given an $\boldsymbol{x}$ of size $256\times256$ pixels and $t=5$, the output of $f_{\boldsymbol{\theta_5}}$ is $8\times8$ pixels.

Given \eqref{eq:proposed_model}, let $T_{\boldsymbol{x}, t}$ be the output tensor produced by $f_{\boldsymbol{\theta_t}}$. 
Then, similarly to~\eqref{eq:cam_reformulated}, we propose to generate a 
class score prediction at each resolution $t$ following
\begin{equation}
    \boldsymbol{\hat{y}}^{(c)}_t = \underbrace{\frac{1}{HW} \sum_{i, j}}_\text{GAP} \underbrace{\left[ \sum_{k \in K} w_{k, c} T_{\boldsymbol{x}, t}(k) \right]
    }_\text{CAM at resolution $t$}
    \label{eq:cam_reformulated_ds}
\end{equation}
and final class scores are obtained as a sum over scores at different resolutions:
\begin{equation}
    \boldsymbol{\hat{y}}^{(c)} = \sum_{t} \boldsymbol{\hat{y}}^{(c)}_t 
    \label{eq:final_scores}
\end{equation}
As indicated by \eqref{eq:cam_reformulated_ds}, prior to generating a class prediction, a CAM at resolution $t$ is produced. This heatmap contains both positive and negative contributions from the input image towards class $c$.
However, for the sake of heatmap clarity, we consider \textit{only} the positive contributions towards each class when generating our CAMs. That is, we want to see what part of the image \textit{contributes} to normality/abnormality, as opposed to what part of the image \textit{does not} contribute to normality/abnormality. Thus, our CAMs are generated following
\begin{equation}
    \boldsymbol{\hat{y}_{CAM_t}}^{(c)} = \left[ \sum_{k \in K} w_{k, c} T_{\boldsymbol{x}, t}(k) \right]^{+}
    \label{eq:cam_positive_contrib}
\end{equation}
where $z^{+} = \max(0, z)$. 
A loss based just on this final score alone would not force the network to produce meaningful CAMs at every resolution level. Therefore, we also propose to deeply supervise the side predictions in our proposed loss:
\begin{equation}
    \mathcal{L} \left (
        \boldsymbol{x}, 
        y, 
        \boldsymbol{\hat{\theta}}, 
        \{ \boldsymbol{\hat{y}}^{(c)}_t \}_{c=1,t=1}^{C, T} 
    \right ) = 
    \mathcal{L}_f \left (\boldsymbol{x}, y, \boldsymbol{\hat{\theta}}, \{ \boldsymbol{\hat{y}}^{(c)}_t \}_{c=1,t=1}^{C, T} \right ) 
    + 
    \sum_t \mathcal{L}_s^t \left (\boldsymbol{x}, y, \boldsymbol{\hat{\theta}}, \{ \boldsymbol{\hat{y}}^{(c)}_t \}_{c=1}^{C} \right )
    \label{eq:ds_loss}
\end{equation}
where $\boldsymbol{x}$ is the input image, $y$ the ground truth class label, $\boldsymbol{\hat{\theta}}$ the network parameters, and $\{ \boldsymbol{\hat{y}}^{(c)}_t \}_{c=1,t=1}^{C, T}$ represent the score predictions for each class $c$ at resolution $t$. Both $\mathcal{L}_f(\cdot)$ and $\mathcal{L}_s^t(\cdot)$ are denoted $\mathcal{L}_f$ and $\mathcal{L}_s^t$ for a simplified notation. $\mathcal{L}_f$ is defined as
\begin{equation}
    \mathcal{L}_f = 
    -y \log\left[\sigma \left(\sum_{t} \boldsymbol{\hat{y}}^{(c)}_t \right)_{c=1} \right] 
    - 
    (1 - y) \log\left[\sigma\left( \sum_{t} \boldsymbol{\hat{y}}^{(c)}_t \right)_{c=0}\right]
    \label{eq:final_loss}
\end{equation}
where $\sigma(\cdot)_{c}$ represents the softmax function for class index $c$. $\mathcal{L}_s^t$ is the side loss for the prediction at each different resolution $t$, defined as:
\begin{equation}
    \mathcal{L}_s^t = 
    -y \log\left[\sigma \left( \boldsymbol{\hat{y}}^{(c)} \right)_{c=1} \right] \\
    - 
    (1 - y) \log\left[\sigma\left( \boldsymbol{\hat{y}}^{(c)} \right)_{c=0}\right]
    \label{eq:side_loss}
\end{equation}
In addition to the network generating CAMs at every resolution prior to generating the scores as part of the prediction pipeline, the combined loss $\mathcal{L}$ proposed allows for the validation of the accuracy at each resolution depth of the network. We refer to the architecture that implements the model in \eqref{eq:proposed_model} with embedded CAMs at different resolutions following \eqref{eq:cam_reformulated_ds} and loss \eqref{eq:ds_loss} as ResNet-18-CAM-DS (see fig~\ref{fig:cars_method}).

\section{Results}
\label{sec:results}
\begin{table*}
	\centering
	\caption{Results for ResNet-18 (baseline model) on frame classification over the testing set of each fold of the IPCL dataset.}
	\begin{tabular}{lcccccc}
		\hline
		\multicolumn{1}{c}{\bfseries Measure (\%)} &
		\multicolumn{1}{c}{\bfseries Fold 1} & 
		\multicolumn{1}{c}{\bfseries Fold 2} & 
		\multicolumn{1}{c}{\bfseries Fold 3} & 
		\multicolumn{1}{c}{\bfseries Fold 4} & 
		\multicolumn{1}{c}{\bfseries Fold 5} & 
		\multicolumn{1}{c}{\bfseries Average} \\
		\hline
		Sensitivity   &  $99.1$   &   $96.6$   &   $96.7$   &   $99.5$   &   $64.3$   &   $91.2$  \\
		Specificity   &  $87.9$   &   $74.7$   &   $62.1$   &   $84.9$   &   $100.0$  &   $81.9$  \\
		Accuracy      &  $94.8$   &   $90.0$   &   $77.2$   &   $92.8$   &   $66.8$   &   $84.3$  \\
		$F_1$ score   &  $95.8$   &   $93.1$   &   $79.0$   &   $93.7$   &   $78.2$   &   $88.0$  \\
	\end{tabular}
	\vspace{0.2cm}
	\label{tab:resnet_18_result_frames}
\end{table*}
\begin{table*}
	\centering
	\caption{Results for ResNet-18-CAM on frame classification over the testing set of each fold of the IPCL dataset.}
	\begin{tabular}{lcccccc}
		\hline
		\multicolumn{1}{c}{\bfseries Measure (\%)} &
		\multicolumn{1}{c}{\bfseries Fold 1} & 
		\multicolumn{1}{c}{\bfseries Fold 2} & 
		\multicolumn{1}{c}{\bfseries Fold 3} & 
		\multicolumn{1}{c}{\bfseries Fold 4} & 
		\multicolumn{1}{c}{\bfseries Fold 5} & 
		\multicolumn{1}{c}{\bfseries Average} \\
		\hline
		Sensitivity   &  $98.6$   &   $94.6$   &   $95.4$   &   $97.6$   &   $75.9$   &   $92.4$  \\
		Specificity   &  $91.7$   &   $89.8$   &   $65.9$   &   $89.4$   &   $100.0$  &   $87.4$  \\
		Accuracy      &  $95.9$   &   $93.1$   &   $78.8$   &   $93.8$   &   $77.6$   &   $87.8$  \\
		$F_1$ score   &  $96.7$   &   $95.0$   &   $79.8$   &   $94.4$   &   $86.3$   &   $90.4$  \\
	\end{tabular}
	\vspace{0.2cm}
	\label{tab:resnet_18_cam_result_frames}
\end{table*}
\begin{table*}
	\centering
	\caption{Results for ResNet-18-CAM-DS on frame classification over the testing set of each fold of the IPCL dataset.}
	\begin{tabular}{lcccccc}
		\hline
		\multicolumn{1}{c}{\bfseries Measure (\%)} &
		\multicolumn{1}{c}{\bfseries Fold 1} & 
		\multicolumn{1}{c}{\bfseries Fold 2} & 
		\multicolumn{1}{c}{\bfseries Fold 3} & 
		\multicolumn{1}{c}{\bfseries Fold 4} & 
		\multicolumn{1}{c}{\bfseries Fold 5} & 
		\multicolumn{1}{c}{\bfseries Average} \\
		\hline
		Sensitivity   &  $99.6$   &   $91.3$   &   $98.3$   &   $98.9$   &   $80.5$   &   $93.7$  \\
		Specificity   &  $81.3$   &   $95.1$   &   $96.6$   &   $89.3$   &   $99.8$   &   $92.4$  \\
		Accuracy      &  $92.5$   &   $92.4$   &   $97.4$   &   $94.5$   &   $81.9$   &   $91.7$  \\
		$F_1$ score   &  $94.1$   &   $94.4$   &   $97.0$   &   $95.1$   &   $89.2$   &   $94.0$  \\
	\end{tabular}
	\vspace{0.2cm}
	\label{tab:resnet_18_cam_ds_result_frames}
\end{table*}
%
%
Our recording protocol (see section \ref{sec:imaged_areas}) enforces that areas recorded in the short patient clips are biopsied. Histopathology labels (normal/abnormal) corresponding to the biopsied specimen are propagated to all the frames of the clip. It is then of interest to evaluate the agreement between the label assigned to each individual frame (based on patient's histopathology) and its correlation to the assessment made by visual inspection of IPCL patterns. A team of $12$ senior clinicians with experience in endoscopic imaging of oesophageal cancer labelled $158$ images from the dataset (randomly picked across patients and manually filtered so that quasi-identical images are not included). A \SI{25}{\percent} per IPCL pattern class (normal, B1, B2, B3) is kept across the sample (leading to an imbalance \SI{25}{\percent} normal, \SI{75}{\percent} abnormal). The inter-rater agreement was evaluated using the Krippendorff's $\alpha$ coefficient, where values \SI{0}{\percent} and \SI{100}{\percent} represent extreme disagreement and perfect agreement respectively, $\alpha >= $ \SI{80}{\percent} indicates reliable agreement, and $\alpha >= $ \SI{66.7}{\percent} tentative agreement \cite{Krippendorff2004}. The Krippendorff's $\alpha$ obtained for the senior clinicians was \SI{76.7}{\percent}. The labels of each clinician were also compared to the 
histopathology, obtaining an average sensitivity, specificity, accuracy, and $F_1$ score (given in \%, with a \SI{95}{\percent} confidence interval) across the $12$ clinicians of $97.0$ [$92.1$, $1.0$], $88.0$ [$49.6$, $1.0$], $94.7$ [$83.9$, $99.7$], and $96.5$ [$89.7$, $99.8$] respectively.

%
%
We report the quantitative classification results for ResNet-18, ResNet-18-CAM, and ResNet-18-CAM-DS in tables \ref{tab:resnet_18_result_frames}, \ref{tab:resnet_18_cam_result_frames}, and \ref{tab:resnet_18_cam_ds_result_frames} respectively. 
ResNet-18-CAM-DS achieved an average sensitivity, specificity, accuracy, and $F_1$ score of \SI{93.7}{\percent}, \SI{92.4}{\percent}, \SI{91.7}{\percent}, and \SI{94.0}{\percent} respectively, all of them better than those achieved by ResNet-18 and ResNet-18-CAM. Accuracy is only three percentage points away from the average of clinical raters. 
Across all folds a total of $60$ patient clips ($12$ per fold) are predicted to be normal/abnormal. The binary class estimation for each clip is computed following \eqref{eq:patient_estimation}. Each patient in the dataset folder has a unique identification number. We will refer to them in this section to facilitate the search of these patients in the dataset folder. Following \eqref{eq:patient_estimation} to estimate the class of a patient clip, ResNet-18 fails on three patients. Folds $1$, $2$, and $4$ fail on patient $158$ (false positive), fold $3$ fails on patient $143$ (false positive), and fold $5$ fails on patient $66$ (false negative). ResNet-18-CAM fails on two patients, $143$ (false positive) on fold 3, and $66$ (false negative) on fold $5$. ResNet-18-CAM-DS fails only on folds $1$ and $4$ in patient $158$ 
(see supplementary material for some frames of these problematic patients).
%
%
%
In fig. \ref{fig:ipcl_cams} a qualitative comparison is shown between the class activation maps produced for the abnormal class by ResNet-18-CAM-DS (at its highest resolution) and the standard class activation maps proposed by Zhou et al. \cite{Zhou2015}.
As our system is designed as a CADe, we have computed the ROC curve (see supplementary material) to inform the consequences that several choices of sensitivity have on specificity. The AUC of the system is \SI{95.8}{\%}.
%
%
%
%
%

\section{Discussion and Conclusion}
\label{sec:discussion}
Our proposed method ResNet-18-CAM-DS achieves slightly higher average accuracy (\SI{91.7}{\percent}) across 
folds than our baseline ResNet-18 (\SI{84.3}{\percent}). Although the automated classification accuracy 
(\SI{91.7}{\percent}) is still below the average achieved by the clinicians (\SI{94.7}{\percent}), it
performs better than some of them (their CI low value is \SI{83.9}{\percent}). It is also encouraging to 
see that accuracy did not decrease at the expense of an improved explainability. More data and further 
methodological refinements will most likely lead to improved accuracy. Qualitative results in 
fig. \ref{fig:ipcl_cams} seem to indicate that the network is \textit{looking at} IPCL patterns to assess 
abnormality, which aligns with the clinical practice. However, we have not observed high activations over 
the large green submucosal vessels in the heatmaps for the normal class. This suggests that they may not be 
used by the network as an aid to solving the classification problem. Future work could concentrate on adding 
an attention mechanism to the network in order to consider such vessels as a feature of normal images.
\section*{Declaration of conflicting interests, ethical approval and informed consent}
R. J. H. has received research grant support from Pentax Medical, Cook Endoscopy, Fractyl Ltd, Beamline Ltd and Covidien plc to support research infrastructure. 
T. V. owns shares from Mauna Kea Technologies, Paris, France. 
The other authors declare that they have no conflict of interest.
%
All procedures performed in studies involving human participants were in accordance with the ethical standards of the institutional and/or national research committee and with the 1964 Helsinki declaration and its later amendments or comparable ethical standards. The Institutional Review Board of E-Da Hospital approved this study (IRB number: EMRP-097-022. July 2017).
%
Informed consent was obtained from all individual participants included in the study.

%


\bibliographystyle{spmpsci}      
\bibliography{library}   



\end{document}